\documentclass[aip, amsmath, amssymb, reprint]{revtex4-1}
\usepackage{graphicx,verbatim}
\usepackage{braket}
\usepackage{color}

\usepackage{soul}
\usepackage[normalem]{ulem}
\usepackage[x11names]{xcolor}

\usepackage{hyperref}
\usepackage[colorinlistoftodos]{todonotes}

\begin{document}

\title{Steady-state properties of multi-orbital systems using quantum Monte Carlo}

\author{A.\ Erpenbeck}
    \affiliation{ 
	Department of Physics, University of Michigan, Ann Arbor, Michigan 48109, USA
	}
	
\author{T. Blommel}
    \affiliation{
	Department of Physics, University of Michigan, Ann Arbor, Michigan 48109, USA
	}    
	
\author{L. Zhang}
    \affiliation{
	Department of Physics, University of Michigan, Ann Arbor, Michigan 48109, USA
	} 
	
\author{W.-T. Lin}
    \affiliation{
	Department of Physics, University of Michigan, Ann Arbor, Michigan 48109, USA
	}

\author{G.\ Cohen}
	\email{gcohen@tau.ac.il}
	\affiliation{
	The Raymond and Beverley Sackler Center for Computational Molecular and Materials Science, Tel Aviv University, Tel Aviv 6997801, Israel
	}
	\affiliation{
	School of Chemistry, Tel Aviv University, Tel Aviv 6997801, Israel
	}
	
\author{E.\ Gull}
	\email{egull@umich.edu}
	\affiliation{
	Department of Physics, University of Michigan, Ann Arbor, Michigan 48109, USA
	}

\date{\today}

\begin{abstract}
    A precise dynamical characterization of quantum impurity models with multiple interacting orbitals is challenging. 
    In quantum Monte Carlo methods, this is embodied by sign problems.
    A dynamical sign problem makes it exponentially difficult to simulate long times. 
    A multi-orbital sign problem generally results in a prohibitive computational cost for systems with multiple impurity degrees of freedom even in static equilibrium calculations.
    Here, we present a numerically exact inchworm method that simultaneously alleviates both sign problems, enabling simulation of multi-orbital systems directly in the equilibrium or nonequilibrium steady-state.
    The method combines ideas from the recently developed steady-state inchworm Monte Carlo framework [Phys.\ Rev.\ Lett.\ 130, 186301 (2023)] with other ideas from the equilibrium multi-orbital inchworm algorithm [Phys.\ Rev.\ Lett.\ 124, 206405 (2020)]. 
    We verify our method by comparison with analytical limits and numerical results from previous methods.
\end{abstract}

\maketitle

\section{Introduction}\label{sec:introduction}

    Quantum impurity models comprise a few strongly interacting orbitals coupled to an extensive noninteracting environment.
    Originally developed in the context of magnetic impurities in materials,\cite{Anderson_Localized_1961}
    they are now commonly used to study topics ranging from quantum phase transitions and non-Fermi liquid behavior\cite{Hewson_Kondo_1997, Vojta_Quantum_2003} 
    to nonequilibrium transport in mesoscopic quantum dots\cite{Datta_Electronic_1997,Cronenwett_Tunable_1998, Goldhaber_Kondo_1998} and molecular electronics.\cite{Nitzan_Electron_2003,Cohen_Greens_2020}
    One of their predominant usages is for embedding theories such as the dynamical mean field theory (DMFT)\cite{Metzner_Correlated_1989,Georges_Dynamical_1996} 
    and its extensions,\cite{Maier_Quantum_2005, Rohringer_Diagrammatic_2018, Aoki_Nonequilibrium_2014, Freericks_Nonequilibrium_2006} 
    as well as self-energy embedding theories.\cite{Kananenka_Systematically_2015, Zgid_Finite_2017}
    These frameworks describe extensive systems like strongly correlated materials by mapping them onto quantum impurity models, which enables a detailed treatment of strong local correlations while considering the remaining system as an environment.\cite{Kotliar_Strongly_2004,Kotliar_Electronic_2006}
    Finding accurate solutions for the underlying generalized impurity model, which often includes multiple orbitals, is a challenging task.   
    Various methods have been used in this context, 
    including exact diagonalization,\cite{Caffarel_Exact_1994, Iskakov_Exact_2018}
    renormalization group techniques,\cite{Bulla_Zero_1999, Bulla_Numerical_2008}
    tensor network representations,\cite{White_Real_2004, Wolf_Chebyshev_2014, Bauernfeind_Fork_2017, Park_Tensor_2024} 
    and hierarchical equations of motion.\cite{tanimura_time_1989, Hou_Hierarchical_2014, tanimura_numerically_2020}
    While successful in numerous scenarios, many of these approaches are constrained to specific parameter or energy ranges, struggle to describe strong correlations, or resort to a rough/specialized description of the environment,
    which limits their applicability.
    Access to dynamical information on the real frequency axis and at both low and high energies, especially in systems driven away from equilibrium, is generally more limited than imaginary time or static information at equilibrium and low energy physics.

    One of the most prevalent methods used in the treatment of multi-orbital impurity models, especially for strongly correlated systems and in the context of quantum embedding methods, is continuous-time quantum Monte Carlo (CTQMC).\cite{Rubtsov_Continuous_2005, Werner_Continuous_2006, Werner_Hybridization_2006, Haule_Quantum_2007, gull_continuous-time_2011}
    These methods use Monte Carlo integration to stochastically sum up perturbative expansions to arbitrary order.
    Different formulations exist, e.g. those based on expansions in the many-body interaction\cite{Rubtsov_Continuous_2004, Rubtsov_Continuous_2005} and the impurity--environment hybridization,\cite{Werner_Continuous_2006, Werner_Hybridization_2006, Haule_Quantum_2007}
    as well as on auxiliary field techniques.\cite{Gull_Continuous_2008, Gull_Submatrix_2011} 
    CTQMC approaches are generally expensive to use when high precision is required, because of the slow convergence of Monte Carlo integration.
    Nevertheless, they excel in the detailed description of a structured environment.
    Furthermore, the computational expense is somewhat compensated for by their reliable accuracy and by the fact that they are extremely well-suited to taking advantage of massively parallel computing resources.
    
    Advancements like bold-line strategies\cite{Prokofev_Bold_2007, Haule_Dynamical_2010, Gull_Bold_2010, Antipov_Voltage_2016} 
    use resummation techniques to achieve accurate and precise results at relatively low orders.
    While each variant has its own advantages and drawbacks,\cite{gull_continuous-time_2011} they are all able to efficiently describe detailed environments, allowing for the handling of general bosonic and fermionic baths.
    However, CTQMC methods encounter challenges known as sign problems that plague all Monte Carlo techniques for quantum systems,\cite{Troyer_Computational_2005} resulting in an exponential growth in computational complexity as a function of certain parameters.
    Sign problems manifest strongly in systems away from certain symmetry points and depend on the particular formulation.\cite{Rubtsov_Continuous_2004, Gull_Continuous_2008}
    For example, the hybridization expansion handles interactions non-perturbatively and is particularly effective in the presence of strong and complex interactions.\cite{Gull_Performance_2007}
    However, it faces multi-orbital sign problems when impurity orbitals can be mixed by their coupling to the environment,\cite{wang_quantum_2010, gorelov_relevance_2009} making it difficult to approach low temperatures even in imaginary time simulations.
    For systems out of equilibrium, which are described by real-time CTHYB techniques,\cite{Muhlbacher_Real_2008, Schiro_Real_2009, Eckstein_Thermalization_2009, Werner_Diagrammatic_2009, Schiro_Real_2010, Werner_Weak_2010, Kubiczek_Exact_2019}  an inherent dynamical sign problem limits simulations to short timescales. 
    This makes an accurate description of nonequilibrium situations extremely difficult, a challenge that has inspired various theoretical approaches aimed at overcoming this limitation.\cite{Mak_Multilevel_1998,Muhlbacher_Crossover_2003,Cohen_Memory_2011,Gull_Numerically_2011,Cohen_Numerically_2013, Cohen_Greens_2014_2,Cohen_Greens_2014, Profumo_Quantum_2015, Bertrand_Quantum_2019, Bertrand_Reconstructing_2019, Moutenet_Cancellation_2019,Kubiczek_Exact_2019,Jeannin_Cross_2024,vanhoecke_diagrammatic_2024,vanhoecke_kondo-zeno_2024, Nestmann_RealTimeTransport_2024}
    
    The inchworm algorithm is one effective strategy for mitigating sign problems.\cite{cohen_taming_2015}
    The inchworm technique combines a CTQMC method with a time-stepping scheme that iteratively propagates the system in time, making optimal use of the information obtained at previous times to make propagation to longer times effectively easier.
    Originally developed for the hybridization expansion in the Anderson impurity model,\cite{cohen_taming_2015} inchworm techniques have since also been applied to two different expansions in the spin--boson model\cite{Chen_Inchworm_2017,Chen_Inchworm_2017_2, goulko_transient_2024} and to spin chains.\cite{wang_quantum_2010}
    Their formulation within the interaction expansion has been applied to fermionic impurity models and lattice models.\cite{li_interaction-expansion_2022}
    The correctness and computational scaling of the method have been assessed mathematically,\cite{cai_inchworm_2023, cai_numerical_2023, cai_bold-thin-bold_2023}
    supported by ongoing algorithmic advancements.\cite{boag_inclusion-exclusion_2018, yang_inclusion-exclusion_2021, cai_fast_2022, Strand_Inchworm_2023}
    Inchworm algorithms have been implemented on the Matsubara contour, the two-branch Keldysh contour, and the three-branch Konstantinov--Perel' contour.
    They have been used to calculate observables such as Green’s functions (GFs) and currents,\cite{Antipov_Currents_2017,Chen_Auxiliary_2019} as well as the full counting statistics of both particle\cite{Ridley_Numerically_2018, pollock_reduced_2022} and energy\cite{Ridley_Numerically_2019} transport.
    This facilitates the exploration of nonequilibrium correlation effects, addressing phenomena such as the voltage splitting of the Kondo resonance,\cite{Krivenko_Dynamics_2019} nonequilibrium full counting statistics in the Kondo regime,\cite{Ridley_Numerically_2018, Ridley_Numerically_2019, Erpenbeck_Revealing_2021} the influence of a structured environment\cite{Ridley_Numerically_2018, Ridley_Numerically_2019, Erpenbeck_Revealing_2021, Erpenbeck_Shaping_2023} and dynamical phases in the spin--boson model.\cite{goulko_transient_2024}
    
    Furthermore, inchworm techniques have featured in studies of extended systems based on DMFT. 
    This includes a proof-of-concept of a real-time equilibrium DMFT calculation,\cite{Dong_Quantum_2017} 
    a study of nonequilibrium switching between phases,\cite{Kleinhenz_Dynamic_2020} 
    an exploration of the structure of the Kondo singlet out of equilibrium,\cite{Erpenbeck_Resolving_2021} 
    studies of localization dynamics in driven many-body systems,\cite{Atanasova_Correlated_2020,Atanasova_Stark_2024} 
    a description of strongly correlated transport in nanostructures,\cite{Erpenbeck_Shaping_2023} 
    simulations of long-lived dynamics in photodoped Mott insulators,\cite{Kunzel_Numerically_2024} 
    and a study of the multi-orbital phases in a bilayer Hubbard model.\cite{goldberger_dynamical_2024}
    
    Despite these successes, an inchworm approach to nonequilibrium/real-time dynamics in multi-orbital systems has so far been elusive.
    While not conceptually difficult, the expected computational cost of a direct calculation using such an algorithm with current implementations would require a prohibitive amount of supercomputer-class resources, giving it relatively limited applicability.
    However, two breakthroughs---which have also enabled several of the latest publications mentioned above---now place this goal within reach.
    First, in Ref.~\onlinecite{Eidelstein_Multiorbital_2020}, the inchworm method was shown to perform well against the multi-orbital sign problem in imaginary-time equilibrium problems.
    Second, in Ref.~\onlinecite{Erpenbeck_Quantum_2023}, an inchworm method was introduced that can directly access the equilibrium or nonequilibrium steady-state without the need for expensive time propagation.
   
    In this publication, we introduce a numerically exact inchworm multi-orbital steady-state (InchMOSS) framework that is based on a combination of the ideas in Ref.~\onlinecite{Eidelstein_Multiorbital_2020} and  Ref.~\onlinecite{Eidelstein_Multiorbital_2020}.
    We benchmark the method against known data in two special but nontrivial limits where analytical or established numerical methods can still be applied, thereby validating the new framework.
    
    The outline of this paper is as follows.
    We review the methodology in Sec.~\ref{sec:method}, where the multi-orbital Anderson model is introduced in Sec.~\ref{sec:model}. 
    In Sec.~\ref{sec:inchworm}, we provide a comprehensive introduction to the inchworm methodology, employing a pedagogical approach that begins with the hybridization expansion and progressively derives the formulation of the steady-state framework.
    The algorithmic details and their implications are discussed in Sec.~\ref{sec:numerics}.
    We provide benchmarks for our multi-orbital inchworm implementation in Sec.~\ref{sec:benchmarks}, where we consider a multi-orbital system that can be decomposed into decoupled orbitals, allowing for a comparison with results from single-orbital methods.
    Sec.~\ref{sec:conclusion} summarizes our work and provides an outlook.

\section{Methodology}\label{sec:method}

    \subsection{The model}\label{sec:model}
        
        We consider a quantum impurity model described by the Hamiltonian
        \begin{eqnarray}
            H &=&  H_{\text{S}} + H_\text{B} + H_{\text{SB}} , 
        \end{eqnarray}
        where $H_\text{S}$ is the Hamiltonian of the impurity system, which consists of multiple interacting orbitals, and $H_\text{B}$ describes the noninteracting bath.
        The impurity and the bath are coupled via $H_{\text{SB}}$.
        For the scope of this work, the impurity is described by a generalized multi-orbital Anderson impurity model,
        \begin{eqnarray}
            H_\text{S} &=& \sum_{i\sigma} \epsilon_{i\sigma} d_{i\sigma}^\dagger d_{i\sigma} 
                    +\sum_{ijkl\atop\sigma\sigma'} U_{ijkl}^{\sigma\sigma'} d_{i\sigma}^\dagger d_{k\sigma'}^\dagger d_{l\sigma'} d_{j\sigma} . \label{eq:H_S}
        \end{eqnarray}
        $d_{i\sigma}^{(\dagger)}$ are the creation/annihilation operators for an electron of spin $\sigma\in\lbrace\uparrow,\downarrow\rbrace$, and $i$ is the orbital index.
        The associated single-particle energy is denoted by $\epsilon_{i\sigma}$. For brevity, we assume that the single-particle part of the impurity Hamiltonian is diagonal in both orbital and spin degrees of freedom, but this is not a requirement of our method.
        The Coulomb interaction between electrons of different orbitals and different spins is encoded in the tensor $
        U_{ijkl}^{\sigma\sigma'}$.
        
        The bath, which can be used to describe an environment or a set of leads coupled to the impurity, consists of noninteracting fermionic orbitals,
        \begin{eqnarray}
            H_\text{B} &=& \sum_l \sum_{k\in l, \sigma} \epsilon_{k\sigma} c_{k\sigma}^\dagger c_{k\sigma} . \label{eq:H_B}
        \end{eqnarray}
        The bath orbitals can be partitioned into subsets $l$, each of which may be characterized by independent thermodynamic parameters like temperature or chemical potential.
        $k$ then labels the orbitals within bath $l$, with associated creation and annihilation operators $c_{k\sigma}^{(\dagger)}$ and single-particle energy $\epsilon_{k\sigma}$.
        
        The coupling between the impurity and the bath is assumed to be of the form
        \begin{eqnarray}
            H_{\text{SB}} &=& \sum_l \sum_{k\in l, \sigma, i} t_{ki\sigma} c_{k\sigma}^\dagger d_{i\sigma} + \text{h.c.}, \label{eq:H_SB}
        \end{eqnarray}
        where $t_{ki\sigma}$ is the scattering amplitude for an electron of spin $\sigma$ in state $i$ on the impurity into state $k$ in bath $l$.
        Given a system--bath coupling in this form, the influence of the bath on the system can be subsumed in the hybridization functions
        \begin{eqnarray}
            \Delta_{ij\sigma}^<(\tau)	&=&	\sum_l \int d\omega\ e^{i\omega\tau} \cdot \Gamma^{l\sigma}_{ij}(\omega) \cdot f_{l\sigma}(\omega), \\
            \Delta_{ij\sigma}^>(\tau)	&=&	\sum_l \int d\omega\ e^{-i\omega\tau} \cdot \Gamma^{l\sigma}_{ij}(\omega) \cdot (1-f_{l\sigma}(\omega)). \label{eq:def_Delta_>}
        \end{eqnarray}
        Here, $f_{l\sigma}$ is the electronic population for spin $\sigma$ in bath $l$,
        and the bath is modelled by the coupling strength function
        $\Gamma^{l\sigma}_{ij}(\omega) = 2\pi \sum_{k\in l} t_{ki\sigma} t_{kj\sigma}^* \delta(\omega-\epsilon_{k\sigma})$. 
        In most applications, the coupling strength function or the hybridization function are used to define the environment, describe an effective bath within embedding schemes such as DMFT,\cite{Georges_Dynamical_1996, Freericks_Nonequilibrium_2006, Aoki_Nonequilibrium_2014} or account for the microscopic details from a lattice surrounding the impurity.\cite{Erpenbeck_Shaping_2023}
        
        A common way to impose nonequilibrium conditions is via the coupling of the impurity to baths with different occupation functions $f_{l\sigma}$.
        For example, a bias voltage is realized by coupling the impurity to baths with different chemical potentials $\mu_{l\sigma}$, 
        a thermal gradient is realized by coupling the impurity  to baths with different temperatures $T_{l\sigma}$, 
        and an optically driven environment can be realized by an appropriate nonequilibrium occupation function.\cite{Kunzel_Numerically_2024}

    \subsection{Steady-state inchworm method}\label{sec:inchworm}
    
            This section offers a pedagogical introduction to the steady-state inchworm scheme introduced in this work, which is based on the hybridization expansion \cite{Werner_Continuous_2006}.
            The latter is perturbative in the coupling between the impurity and the bath.
            We begin by introducing the main components of the standard inchworm methodology and hybridization expansion, and then derive the corresponding steady-state inchworm framework.
            We assume the reader to be familiar with Keldysh technique.\cite{Haug_Quantum_2008, stefanucci2013nonequilibrium, Kamenev_Field_2023}
            
        \subsubsection{Restricted propagators}\label{sec:propagators}
            The central object of the method is the restricted atomic state propagator,
            \begin{eqnarray}
                \varphi_\alpha^\beta(t)	=	\text{Tr}_\text{B} \left\lbrace \rho_\text{B}
								\bra{\beta}e^{-iHt}\ket{\alpha} \right\rbrace . \label{eq:propagator_single}
            \end{eqnarray}
            We also define a two-time restricted propagator spanning both branches of the Keldysh contour,
            \begin{eqnarray}
                \Phi_\alpha^{\beta'\beta}(t',t)	=	\text{Tr}_\text{B} \left\lbrace \rho_\text{B}
								\bra{\alpha}e^{iHt'}\ket{\beta'} \bra{\beta}e^{-iHt}\ket{\alpha}
								\right\rbrace ,
								\label{eq:propagator_double}
            \end{eqnarray}
            where $t'$ and $t$ are times on different Keldysh branches.
            Here, $\text{Tr}_\text{B}$ is the trace over the bath degrees of freedom, and $\alpha$ and $\beta$ are atomic states in the Fock space of $H_{\text{S}}$.
            In particular, $\alpha$ is the initial state of the impurity (assuming a factorized initial condition between the impurity and bath).
            $\rho_\text{B}$ is the initial density matrix of the bath.
            
            The two-branch restricted propagator $\Phi_\alpha^{\beta'\beta}(t',t)$ is directly related to single-time physical observabes such as the population or the current flowing through the system (cf.\ Sec.~\ref{sec:observabes}).
            The single-branch propagator is important for calculating two-time observables such as the GF and serves as an auxiliary object for calculating the two-branch restricted propagator (cf.\ Sec.~\ref{sec:Keldysh}).
            A discussion of these propagators is given in Refs.~\onlinecite{Cohen_Greens_2014, Erpenbeck_Resolving_2021}.

        \subsubsection{Hybridization expansion and CTQMC} \label{sec:hybridization_expansion}

            We introduce the methodology using the simplest object to which it can be applied, the single-branch propagator $\varphi_\alpha^\beta(t)$.
            Expanding Eq.~(\ref{eq:propagator_single}) in the impurity--bath coupling yields
            \begin{eqnarray}
                \varphi_\alpha^\beta(t)
                &=&
                \sum_{n=0}^\infty (-i)^n 
                            \int_0^t d\tau_1 \dots  \int_0^{\tau_{n-1}} \hspace*{-0.4cm} d\tau_n 
                            \label{eq:propagator_single_expansion}
                            \times \\ && \times
                        \text{Tr}_B \Big\lbrace \rho_B
                            \bra{\beta}
                            e^{-iH_0t}H_{\text{SB}}(\tau_1) \dots H_{\text{SB}}(\tau_n)\ket{\alpha}
                        \Big\rbrace
                        \nonumber
                        ,
            \end{eqnarray}
            where $H_{\text{SB}}(\tau) = e^{iH_0\tau}H_{\text{SB}}e^{-iH_0\tau}$ for $H_0 = H_\text{S} + H_\text{B}$.
            This $H_{\text{SB}}(\tau)$ is the impurity--bath coupling in the interaction picture with respect to $H_0$.
            Eq.~(\ref{eq:propagator_single_expansion}) is formally exact and sums up all orders $n$ of the hybridization expansion, whereby the contribution of every order is given by an $n$-dimensional integral.

\begin{figure}
                \includegraphics{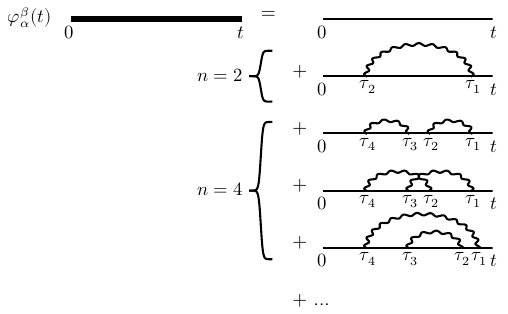}
                \caption{
                	Feynman diagram representation of the hybridization expansion of the single-branch restricted propagator $\varphi_\alpha^\beta(t)$ according to Eq.~(\ref{eq:propagator_single_expansion}). 
                	The figure shows examples of Feynman diagrams up to order $n=4$.
                	Thick lines represent full propagators and thin lines are bare propagators generated by $H_{\text{SB}}=0$. 
                	Wiggly lines represent hybridization lines, which connect two hybridization events at times $\tau_i$.
                	The text describes how specific Feynman diagram are evaluated.
                	}
                \label{fig:feynman_QMC}
            \end{figure}

            The integrands in Eq.~(\ref{eq:propagator_single_expansion}) can be visualized by Feynman diagrams, as illustrated in Fig.~\ref{fig:feynman_QMC}.
            A Feynman diagram of order $n$ contains $n$ hybridization times $\tau_i$ with $i \in \lbrace 1\dots n \rbrace$.
            As $H_{\text{SB}}$ is linear in the creation/annihilation operators of the bath, particle conservation implies that each bath operator $c_{k\sigma}$ at time $\tau_i$ must be matched with its counterpart $c_{k\sigma}^\dagger$ at time $\tau_j$ (where $i \neq j$) to produce a nonzero contribution.
            This is represented by Feynman diagrams where all hybridization times are paired up and connected by
            hybridization lines.
            Hence, only even hybridization orders contribute.
            By using Wick's theorem within the noninteracting bath, all possibilities of connecting a set of hybridization times can be expressed in terms of a determinant, enabling inexpensive evaluation even at high orders.
            Nevertheless, individual Feynman diagrams are not only valuable for compact visualization of contributions to the hybridization expansion, but also play an important role when developing resummation schemes.
            
            For our purposes, Feynman diagrams are composed of thick lines representing ``full'' or ``bold'' propagators referring to $\varphi_\alpha^\beta(t)$, wiggly lines representing hybridizations, and thin lines representing ``bare'' propagators.
            The latter are propagators within the impurity subspace and in the limit $H_{\text{SB}}\rightarrow 0$, given by $\braket{\beta | e^{-iH_{\text{S}}t} | \alpha}$; these are calculated by exact diagonalization of $H_\text{S}$.
            A Feynman diagram of order $2n$ contains $2n$ hybridization times $\tau_i$.
            At each such time, a vertex is present, containing a term from $H_\text{SB}$ with one of $n$ creation operators $d_{j_i\sigma_i}^{\dagger}$, where $i$ is in a subset comprising half the index values; or one of $n$ annihilation operators $d_{j_i\sigma_i}$ with $i$ in the complementary subset of indices.
            Any particular particular Feynman diagram can then be evaluated as the product of a propagator matrix in the impurity subspace, a hybridization contribution, and a sign.
            The propagator matrix is a time-ordered product of all propagators between hybridization times and the creation and annihilation operators, and is given by
            $\sum_{\gamma_1 \dots \gamma_{2n}}
            \braket{\beta | e^{-iH_{\text{S}}(t - \tau_1)} | \gamma_{1}}
            \cdot
            \braket{\gamma_{1} | d_{j_1\sigma_1}^{\dagger_1} | \gamma_2}
            \cdot
            \braket{\gamma_2 | e^{-iH_{\text{S}}(\tau_1 - \tau_2)} | \gamma_{3}}
            \cdot\dots\cdot
            \braket{\gamma_{2n-2} | e^{-iH_{\text{S}}(\tau_{n-1} - \tau_n)} | \gamma_{2n-1}}
            \cdot
            \braket{\gamma_{2n-1} | d_{j_n\sigma_n}^{\dagger_n} | \gamma_{2n}}
            \cdot
            \braket{\gamma_{2n} | e^{-iH_{\text{S}}\tau_n} | \alpha}$.
            The hybridization part is the product of all hybridizations connecting pairs of hybridization times $(\tau_i, \tau_j)$ with $\tau_i>\tau_j$, 
            $\Pi_{(\tau_i, \tau_j)} \Delta_{ij\sigma}^\lessgtr(\tau_i-\tau_j)$, whereby the lesser(greater) component is chosen when an annihilation(creation) operator acts at time $\tau_i$. 
            The sign is given by $-1^{n_{\text{cross}}}$, where $n_{\text{cross}}$ is the number of times that the hybridization lines intersect within the Feynman diagram.
            These rules are then generalized in later sections, where bare propagators between hybridization times are replaced by their bold counterparts.

            Eq.~(\ref{eq:propagator_single_expansion}) can be evaluated using CTQMC techniques.
            A Metropolis Monte Carlo algorithm is used to add or remove hybridization vertices such that all relevant orders are sampled, allowing for efficient evaluation of the hybridization expansion in a wide variety of models and physical regimes (see Ref.~\onlinecite{gull_continuous-time_2011} and references therein for details).
            In real-time simulations, however, this is only feasible for short times $t$.
            The oscillatory nature of the integrands in Eq.~(\ref{eq:propagator_single_expansion}) result in a dynamical sign problem, causing the computational cost for to grow exponentially with $t$.
            Apart from the dynamical sign problem, mulit-orbital impurities are prone to multi-orbital sign problems.
            This sign problem, which is already present in equilibrium formulations on the imaginary time axis, arises when the impurity orbitals are coupled by the baths, i.e.~when the hybridization functions are off-diagonal,\cite{wang_quantum_2010, gorelov_relevance_2009} 
            which is due to $\Delta_{ij\sigma}^\lessgtr$ not being necessarily strictly positive for $i\neq j$.

        \subsubsection{Basics of the inchworm framework} \label{sec:inchworm_scheme}
        
            The inchworm expansion is an efficient way to apply resummation to diagrammatic quantum Monte Carlo calculations, and can be used to suppress both the dynamical sign problem\cite{cohen_taming_2015, Antipov_Currents_2017, Chen_Inchworm_2017, Chen_Inchworm_2017_2, cai_inchworm_2023, cai_numerical_2023}
            and the multi-orbital sign problem.\cite{Eidelstein_Multiorbital_2020,li_interaction-expansion_2022,goldberger_dynamical_2024, Strand_Inchworm_2023}
            Instead of calculating the restricted propagator for a given time $t$ directly, the inchworm approach assumes that the restricted propagator is known up to a time $T<t$, and from there constructs a more efficient expansion for the propagator $\varphi_\alpha^\beta(t)$.
            The latter expansion leverages all the information contained in the restricted propagator at shorter times.
            In mathematical terms, the inchworm methodology provides a map
            $F_{\text{inch}}^\varphi : \lbrace \varphi_{\mu}^{\nu}(\tau) \ | \ \tau \leq T;\ \forall \mu,\nu \rbrace \rightarrow \varphi_\alpha^\beta(t)$.
            The mapping is constructed from the hybridization expansion, but using fully dressed propagators within the temporal region where they are available.

\begin{figure}
                \includegraphics{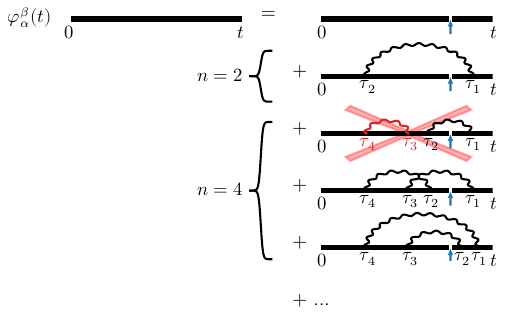}
                \caption{
                    Feynman diagram representation of the inchworm expansion of the single-branch restricted propagator $\varphi_\alpha^\beta(t)$ as given in Eq.~(\ref{eq:Feynman_inch_single_2}). 
                    The blue arrow indicates the time $T$.
                    Red hybridization lines are already included in propagators for shorter times---the corresponding diagrams are therefore not inchworm proper and hence not part of the inchworm expansion.
                    }
                \label{fig:Feynman_inch_single}
            \end{figure}

            In App.~\ref{app:single_branch}, we derive the lowest order expressions for the restricted propagator from the hybridization expansion.
            While Eq.~(\ref{eq:Feynman_inch_single_2}) explicitly provides the lowest two orders of the inchworm expansion, higher-order expressions require specific consideration pertaining the connections between hybridization times, which leads to lengthy equations.
            Feynman diagrams provide a more compact way to visualize terms in the inchworm expansion.
            This is exemplified in Fig.~\ref{fig:Feynman_inch_single}, where the first two lines on the right correspond to the terms given in Eq.~(\ref{eq:Feynman_inch_single_2}). 
            In an inchworm step, which aims to extend the propagator from times up to $T$ to longer times, any Feynman diagram within this expansion must obey a simple rule:
            every set of hybridization lines connected by crossing must contain at least one hybridization event in the interval $(0,T)$, and at least one in the interval $(T, t)$.
            Diagrams obeying this restriction are referred to as ``inchworm proper''.\cite{cohen_taming_2015, Chen_Inchworm_2017, Chen_Inchworm_2017_2}
            Conversely, any cluster of hybridization lines exclusively located at times either larger or smaller than $T$ is already accounted for by the restricted propagators at shorter time intervals and must therefore be excluded from the inchworm expansion. 
            An example of an improper diagram is given in the third line of Fig.~\ref{fig:Feynman_inch_single}.
            
            Because each diagram in the inchworm expansion contains an infinite set of bare diagrams, the expansion converges significantly more rapidly than its bare counterpart.
            However, this efficiency comes with increased computational complexity due to the need to explicitly sum only inchworm proper diagrams.
            Unlike the bare hybridization expansion, inchworm proper diagrams cannot be summed by a determinant. They can be individually enumerated by using Heap's algorithm to sum over permutations, but more efficient algorithms have been proposed.\cite{boag_inclusion-exclusion_2018,yang_inclusion-exclusion_2021,cai_fast_2022}

            Another important aspect of the algorithm involves storing and evaluating the restricted propagator as a function of time. 
            While in the bare hybridization expansion the bare restricted propagator can be easily obtained for any time $t$, the inchworm expansion relies on propagators evaluated at previous times. 
            In practice, an inchworm calculation begins by obtaining the restricted propagator for a sequence of short times using the bare hybridization expansion, where this is feasible.
            Subsequently, the time is incrementally increased by small steps of length $t_{\text{inch}}=t-T$.
            Here, $t_{\text{inch}}$ is a numerical parameter determining various aspects, including the convergence of the expansion with respect to the order of inchworm diagrams to be considered.
            Propagators are stored on the resulting time grid and evaluated by interpolation (linear in current implementations).
            This iterative process, which gradually advances time until reaching the desired time $t$, inspired the name of the algorithm, and results in a linear computational scaling with the number of time steps taken to reach $t$.

            After introducing the fundamentals of the inchworm method, we comment on its similarities and differences with other common approaches based on similar principles. 
            While the inchworm algorithm employs a time-stepping scheme, it is not derived from an expression for the time derivative that is then used for propagation, as is the case for i.e.~the Kadanoff–Baym framework.\cite{kadanoff_quantum, Lipavsky_Generalized_1986, stefanucci2013nonequilibrium} 
            Methods based on time derivatives are typically accurate only for infinitesimal time steps, whereas the inchworm method maintains accuracy over a wide range of inchworm step sizes, significantly enhancing its stability and numerical efficiency. 
            However, larger inchworm steps usually require higher orders for converged results.
            Moreover, the inchworm method is a resummation scheme.
            However, in contrast to self-energy-based resummation schemes, which often rely on high-order self-consistent relations between propagators and the self-energy and which can be challenging to treat numerically, the inchworm method utilizes propagators for previous times. As such, each inchworm step serves as a small correction to propagators at shorter times, thereby improving its numerical feasibility.

        \subsubsection{Inchworm propagation on the Keldysh contour} \label{sec:Keldysh}
        
            Having introduced the inchworm approach for the restricted propagator on a single time branch, we extend the inchworm methodology to the restricted propagator on two branches of the Keldysh contour as defined in Eq.~(\ref{eq:propagator_double}).
            The overall inchworm framework is identical for single- and two-branch restricted propagators, and the inchworm methodology can in principle be formulated using contour times only, such that both propagators are treated on the same footing.
            Here, we distinguish between single- and two-branch restricted propagators for convenience, which leads to two subtle yet important differences.
            First, $\Phi_\alpha^{\beta'\beta}(t',t)$ has two time arguments and an inchworm expansion that extends the propagator from shorter to longer times can be formulated for either time argument; essentially, one can ``inch'' both forwards and backwards in time. 
            Second, as the two-branch restricted propagators incorporates both branches of the Keldysh contour, the inchworm resummation scheme depends not only on information from restricted propagators spanning both contours, but also on those confined to a single contour.
            Accordingly, the map for the two-branch propagator is
            $F_{\text{inch}}^\Phi : 
            \lbrace \Phi_\mu^{\nu'\nu}(\tau',\tau); \varphi_{\mu}^{\nu}(\tau) \ | \ \tau \leq T; \ \tau'\leq t'; \forall \mu,\nu,\nu' \rbrace \rightarrow \Phi_\alpha^{\beta'\beta}(t',t)$.

\begin{figure}[tb!]
                \includegraphics{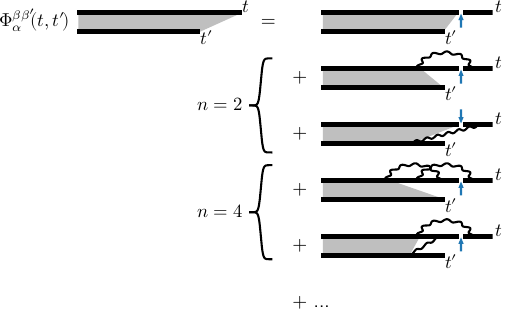}
                \caption{Feynman diagram representation of the inchworm expansion of the two-branch restricted propagator $\Phi_\alpha^{\beta'\beta}(t',t)$ as given in Eq.~(\ref{eq:Feynman_inch_double_2}).
                The blue arrow indicates the time $T$, only inchworm proper diagrams are shown.
                Thicks line represents full single-branch propagators, grey shaded areas that connect the two branches indicate full two-branch propagators. 
                Wiggly lines represent hybridizations.}
                \label{fig:Feynman_inch_double}
            \end{figure}

            In App.~\ref{app:double_branch}, 
            we derive the lowest order expressions for $\Phi_\alpha^{\beta'\beta}(t',t)$ for the second time argument using the hybridization expansion.
            The procedure is equivalent for the first time argument.
            Eq.~(\ref{eq:Feynman_inch_double_2}) explicitly provides the lowest two orders of the inchworm expansion. 
            As for the single-branch case, higher-order expressions become lengthy due to the combinatorics of hybridization lines, 
            and Feynman diagrams offer a more efficient representation of the inchworm scheme, as exemplified in Fig.~\ref{fig:Feynman_inch_double}.
            The first three lines on the right of Fig.~\ref{fig:Feynman_inch_double} correspond to the terms given in Eq.~(\ref{eq:Feynman_inch_double_2}). 
            Notice that the second order includes two contributions, reflecting the possibility of the two hybridization times being located on the same or different Keldysh branches. 
            Despite this distinction and the incorporation of both single-branch and two-branch restricted propagators into the expression, the concept of inchworm proper Feynman diagrams remains unchanged.

            Calculating the two-branch propagator using the inchworm framework involves incrementally advancing the two time arguments by a time step of size $t_{\text{inch}}$ until the desired final times are reached. 
            Consequently, the numerical cost of this algorithm scales at least quadratically with the number of time steps necessary to reach the final time.\cite{cohen_taming_2015, Chen_Inchworm_2017} 
            This type of inchworm method is therefore predominantly limited to intermediate timescales.

        \subsubsection{Formulation directly in the steady-state} \label{sec:steady-state}
        
            Using the inchworm framework, we can compute steady-state properties without the need for propagation, as recently proposed in Ref.~\onlinecite{Erpenbeck_Quantum_2023}.
            This approach relies on two key assumptions:
            \begin{enumerate}
            	\item The steady-state is independent of the impurity’s initial condition $\alpha$.
            	\item The steady-state two-branch propagator only depends on the relative time $\Delta t=t'-t$ rather than the explicit times $t'$ and $t$. 
            \end{enumerate}
            These assumptions motivate an ansatz for the two-branch propagator in the steady-state, $\Phi_{\text{SS}}^{\beta'\beta}(\Delta t)$.
            Replacing $\Phi_\alpha^{\beta'\beta}(t',t) \rightarrow \Phi_{\text{SS}}^{\beta'\beta}(\Delta t)$ in the map $F_{\text{inch}}^\Phi$ derived in Sec.~\ref{sec:Keldysh}
            yields a map that self-consistently relates the steady-state propagator to itself,
            $F_{\text{inch}}^\Phi : 
            \lbrace \Phi_{\text{SS}}^{\mu\nu}(\Delta \tau); \varphi_{\mu}^{\nu}(\tau) \ | \forall \tau,\Delta\tau; \forall \mu,\nu \rbrace \rightarrow \Phi_{\text{SS}}^{\beta'\beta}(\Delta t)$.
            As the steady-state propagator only depends on the relative time, the notion of ``previous times'' is no longer applicable.
            Thus, the steady-state propagator must be known for any given $\Delta t$ to construct the map $F_{\text{inch}}^\Phi$.
            All other elements of the inchworm framework, such as the definition of inchworm proper diagrams, remain unaffected by the steady-state formulation.

            Since $F_{\text{inch}}^\Phi$ maps $\Phi_{\text{SS}}^{\beta'\beta}(\Delta t)$ self-consistently onto itself, the steady-state propagator can be calculated directly without the need for propagation. 
            Conceptually, the self-consistency condition enables us to assert that within the steady-state, advancing one inchworm step forward in time does not alter the steady-state restricted propagator.
            Algorithmically, we start from an initial guess for the steady-state restricted propagator, which is iteratively improved using the map $F_{\text{inch}}^\Phi$ until convergence is achieved. 
            This approach is particularly advantageous over propagation for systems exhibiting long-lived transient behavior. 
            Additionally, since the steady-state restricted propagator only depends on the relative time rather than two explicit times, the steady-state approach scales linearly with the coherence time $t_{\text{max}}$, which we define as the numerical cutoff time at which the steady-state propagator can effectively be assumed to be zero ($\Phi_{\text{SS}}^{\beta'\beta}(\Delta t)\approx 0$ for $\Delta t >t_{\text{max}}$).
            We note that because the self-consistency condition is homogeneous, any self-consistent solution multiplied by a constant also constitutes a solution to the self-consistency cycle. 
            Therefore, we enforce an additional physical normalization condition $\sum_\beta \Phi_{\text{SS}}^{\beta\beta}(0) = 1$ at each iteration.
            This is equivalent to forcing the trace of the density matrix to 1, thereby imposing the conservation of probability.

        \subsubsection{Green's functions and other observables of interest} \label{sec:observabes}
            We wrap up the theory section by showing how certain commonly used physical observables can be evaluated given knowledge of the restricted propagators.
            
            An observable directly related to the two-time restricted propagator is the probability for finding the impurity in state  $\beta$ at time $t$, which is given by
            \begin{eqnarray}
                p_\beta(t) &=&  \Phi_\alpha^{\beta\beta}(t,t) .
            \end{eqnarray}
            Similarly, any observable within the system subspace, given that the impurity began in state $\alpha$ at time $t=0$, can be expressed as
            \begin{eqnarray}
            	\left\langle \hat{O}_{\text{S}}\left(t\right)\right\rangle &=&\sum_{\beta\beta'}\Phi_{\alpha}^{\beta'\beta}\left(t,t\right)\left\langle \beta'\right|\hat{O}_{\text{S}}\left|\beta\right\rangle .
            \end{eqnarray}
            In both cases, the steady-state response, which is obtained at the limit $t\rightarrow\infty$, is given by replacing $ \Phi_\alpha^{\beta'\beta}(t,t)$ with $\Phi_{\text{SS}}^{\beta'\beta}(0)$.
            
\begin{figure}[tb!]
                \includegraphics{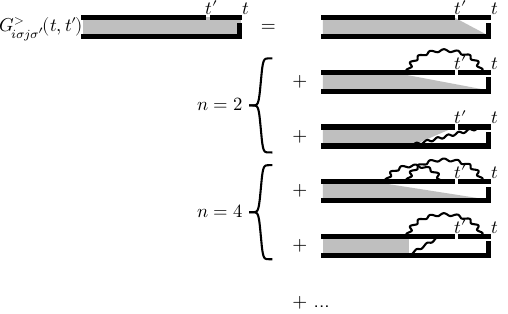}
                \caption{Feynman diagram representation for calculating the greater GF $G^>_{i\sigma j\sigma'}(t,t')$ from the single- and two-branch restricted propagators as given in Eq.~(\ref{eq:Feynman_inch_GF}).
                Notice the similarity to the expansion of the two-time restricted propagator in Fig.~\ref{fig:Feynman_inch_double}, where for the GF, the time $T$ is replaced by the time $t'$.
                As before, thick line are full single-branch propagators, grey shaded areas are full two-branch propagators, while wiggly lines represent hybridizations.}
                \label{fig:Feynman_inch_GF}
            \end{figure}

            The main observabes of this work, however, are the lesser, greater, and retarded GFs, defined in the steady-state as\cite{Haug_Quantum_2008, stefanucci2013nonequilibrium}
            \begin{eqnarray}
                G^<_{i\sigma j\sigma'}(t-t') &=&    i \braket{ d^\dagger_{j\sigma'}(t') d_{i\sigma}(t) } , \\
                G^>_{i\sigma j\sigma'}(t-t') &=&   -i \braket{ d_{i\sigma}(t) d^\dagger_{j\sigma'}(t') } ,
                \label{eq:GF_gtr_def}\\
                G^r_{i\sigma j\sigma'}(t) &=&    \Theta(t) \left(G^>_{i\sigma j\sigma'}(t) - G^<_{i\sigma j\sigma'}(t) \right) ,
            \end{eqnarray}
            where $\Theta$ is the Heaviside step-function.
            These expressions for the GFs can be used as the starting point for the hybridization expansion.
            This procedure makes it possible to express them in terms of single- and two-branch restricted propagators.
            This is demonstrated in App.~\ref{app:GF}, where Eq.~(\ref{eq:Feynman_inch_GF}) is the expression for the lowest two orders for the greater GF in terms of restricted propagators.
            The corresponding Feynman diagram representation is given in Fig.~\ref{fig:Feynman_inch_GF}.
            
            Expressing the GFs in terms of restricted propagators leverages the resummation scheme employed in the inchworm expansion itself. 
            Consequently, the expressions for the GF closely resemble those for the two-branch restricted propagator (compare Eqs.~(\ref{eq:Feynman_inch_double_2}) and (\ref{eq:Feynman_inch_GF}) and their respective Feynman diagram representations in Figs.~\ref{fig:Feynman_inch_double} and \ref{fig:Feynman_inch_GF}).
            In particular, associating the time $t'$ in the GF with time $T$ in the inchworm framework and incorporating the respective fermionic creation and annihilation operators at $t'$ and $t$, GF calculations are equivalent to those of a two-branch restricted propagator. 
            This similarity arises from the shared underlying resummation structure and while GF calculations can be expressed using inchworm proper diagrammatics, they do not employ an inching or time-stepping scheme for computation.

            Note that time-dependent and steady-state calculations only differ in the usage of the two-branch restricted propagator or its steady-state counterpart.
            In the steady-state scenario, GFs in the energy representation are derived from their time-dependent counterparts using the relation
            $G^{\lessgtr/r}_{i\sigma j\sigma'}(\omega) = \int G^{\lessgtr/r}_{i\sigma j\sigma'}(t-t') e^{i\omega(t-t')} d\omega$.
            The spectral function is then defined as
            $A_{i\sigma j\sigma'}(\omega) = -\frac{1}{\pi}\text{Im}\lbrace G^{r}_{i\sigma j\sigma'}(\omega) \rbrace$.

    \subsection{Algorithm and scaling analysis}\label{sec:numerics}
    
        After outlining the theory of the inchworm methodology and its formulation in the steady-state, we describe how an InchMOSS calculation is structured while commenting on numerical scaling and implementational aspects. 
        
        \subsubsection{Phases of a steady-state calculation}
        
            Our steady-state inchworm calculation consists of three consecutive phases:
            \begin{enumerate}
                \item 
                Calculation of the single-branch restricted propagator $\varphi_\alpha^\beta(t)$:
                Using the map $F_{\text{inch}}^\varphi$, we compute $\varphi_\alpha^\beta(t)$ employing an inchworm propagation scheme.
                Here, the time values $t$ are incrementally increased by a small time-step of length $t_\text{inch}$ until reaching the final time $t_{\text{max}}$. 
                The scaling is linear with respect to $t_{\text{max}}$ for fixed $t_\text{inch}$.
            
                \item 
                Calculation of the two-branch restricted propagator in the steady-state $\Phi_{\text{SS}}^{\beta'\beta}(\Delta t)$:
                Using the map $F_{\text{inch}}^\Phi$, we compute $\Phi_{\text{SS}}^{\beta'\beta}(\Delta t)$, which requires knowledge of  $\varphi_\alpha^\beta(t)$ from the previous phase. 
                While the map $F_{\text{inch}}^\Phi$ was derived from the inchworm scheme, the computation of $\Phi_{\text{SS}}^{\beta'\beta}(\Delta t)$ is based on iteration until self-consistency rather than inchworm propagation. 
                The scaling is linear with the number of time-points used to represent $\Phi_{\text{SS}}^{\beta'\beta}(\Delta t)$, implying a linear scaling with $t_{\text{max}}$. 
                Further, this phase scales linearly with the number of iterations, typically ranging from $100-200$ for a single orbital, with a tendency to increase with the number of orbitals.
                
                \item 
                Calculation of the GFs $G^{\lessgtr/r}_{i\sigma j\sigma'}(\Delta t)$:
                This phase relies on the knowledge of $\varphi_\alpha^\beta(t)$ and $\Phi_{\text{SS}}^{\beta'\beta}(\Delta t)$. 
                While inchworm proper diagrammatics is leveraged, the calculation of the GFs doesn't entail inchworm propagation. The scaling is linear with the number of time-points used to represent the GFs.
            \end{enumerate}

        \subsubsection{Details on the implementation}
        
            Our methodology requires the computation of high-dimensional integrals. 
            Despite recent advancements in techniques using tensor approximations\cite{nunez_learning_2022, erpenbeck_tensor_2023} or low-discrepancy sequences,\cite{Macek_Quantum_2020, Bertrand_Quantum_2021, Strand_Inchworm_2023} which have shown improved numerical efficiency, we use a Metropolis Monte Carlo algorithm\cite{landau_guide_2021, krauth_statistical_2006, gubernatis_quantum_2016} to calculate these integrals. 
            This algorithm converges to the correct result with a rate of $1/\sqrt{N}$, where $N$ is the number of Monte Carlo samples.
            The relatively slow convergence of the Metropolis method is compensated for by flexible importance sampling and a high level of reliability as long as ergodicity is achieved.
            
            To enable a general multi-orbital description, our approach represents all restricted propagators as matrices in the basis of the impurity states. 
            We employ the $\lVert \cdot \rVert_1$ matrix norm to determine the importance sampling weights within the Metropolis Monte Carlo algorithm. 
            Alternative matrix norms also prove efficient, 
            with the optimal norm depending on the system under consideration.
            Employing a matrix representation reduces concatenating restricted propagators and calculating bare propagators
            to matrix multiplications and exponential function computations of matrices. 
            In the most general scenario, this scales as $\mathcal{O}(n_\text{states}^3)$, 
            where $n_\text{states}=2^{n_\text{orb}}$ represents the dimension of the impurity Fock space and $n_\text{orb}$ is the number of spin-orbitals of the impurity. 
            Efficiency can be enhanced by exploiting system symmetries or confining the analysis to the single-orbital case, where various simplifications, such as the utilization of a ``segment picture", become possible.\cite{Werner_Continuous_2006, gull_continuous-time_2011}
            The extension of these concepts to multi-orbital inchworm remains a topic for future research.
    
            We investigated various grids for representing restricted propagators. While employing Chebyshev nodes can offer advantages and reduce numerical costs in specific cases, we observed that an equidistant grid generally yields the most reliable results. 
            Additionally, we explored convergence acceleration techniques such as DIIS\cite{pulay_convergence_1980, pulay_improved_1982, Shepard_Some_2007} for the iterative calculation of the steady-state two-branch propagator but did not observe any consistent improvement in convergence behavior.
            Finally, we remark that we use the inclusion--exclusion algorithm to compute inchworm proper diagrams, which scales as $\mathcal{O}(\gamma^n)$, with $\gamma\approx1.33$ and $n$ the inchworm order.\cite{boag_inclusion-exclusion_2018}

\section{Results}\label{sec:benchmarks}

    In this results section, we showcase the feasibility of the InchMOSS method to accurately describe multi-orbital quantum impurity models. 
    Thereby, we distinguish between cases where the orbitals effectively decouple and a description in terms of single orbital methods is in principle straightforward,
    and cases where the orbitals of the impurity are coupled via interactions on the impurity itself and via the coupling to the baths.

    \subsection{Interacting and non-interacting physics in decoupled orbitals}\label{sec:benchmarks_1}
        \begin{figure}
            \includegraphics{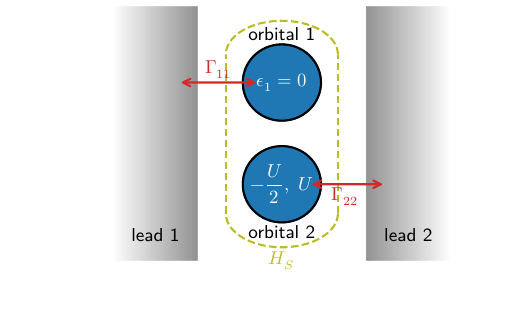}
            \includegraphics{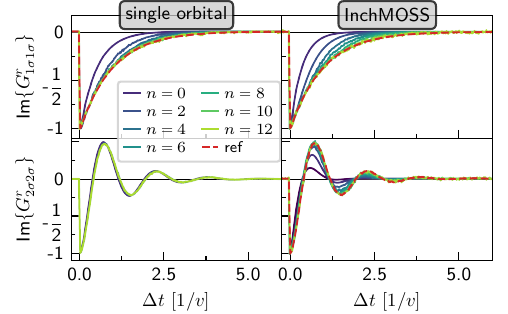}
            \caption{First benchmark for a two-orbital system.
                     Top: Sketch of the system which can be decomposed into two decoupled orbitals.
                     Blue circles depict the impurity orbitals, gray regions represent the leads, and red arrows indicate the presence of impurity--lead hybridization.
                     Bottom: Convergence of the retarded GF with inchworm order $n$ for the first orbital (top panels) and the second orbital (bottom panels).
                     Results from the separate solution of the two single-orbital problems are presented in the left panels, and results from the multi-orbital implementation treating both orbitals simultaneously are presented in the right panels.
                     The system parameters are $\epsilon_1=0$, $U= -2\epsilon_2 = 8v$.
                     The leads are described in the wide-band limit with a smooth cutoff with maximum coupling strength $\Gamma_{11} = 2\Gamma_{22} = v^2$.
                     The inverse temperature is $\beta=0.1/v$.
                     The reference in each case is given by a red dashed line.
                     An analytic result is used for orbital one, and the converged numerical result from the single-orbital calculation ($n=12$ case in left bottom panel) is used as the reference for the multi-orbital treatment of orbital two (right bottom panel).
                     }
            \label{fig:benchmark1}
        \end{figure}
        
        With our first benchmark, we establish that 
        (i) the InchMOSS formalism is consistent with the previous single-orbital framework from Ref.~\onlinecite{Erpenbeck_Quantum_2023}, 
        (ii) is capable of reproducing analytic results as well as numerical results from other methods on the same footing, and
        (iii) recovers the symmetries in multi-orbital systems.
        Moreover, we use this benchmark to showcase the connection between convergence behavior with respect to inchworm order and the number of orbitals.

        We consider the spinfull two-orbital Anderson impurity model, described by the Hamiltonian
        \begin{eqnarray}
            H_\text{S} &=& \epsilon_1 \sum_{\sigma} n_{1\sigma}
                    -\frac{U}{2} \sum_{\sigma}  n_{2\sigma} + U n_{2\uparrow}n_{2\downarrow} .                
        \end{eqnarray}
        The first orbital $i=1$ is noninteracting, and without loss of generality, we set $\epsilon_1=0$.
        The second orbital $i=2$ is interacting and particle-hole symmetric, with $U=8v$. 
        Our unit of energy is $v$, which is related to the impurity-bath coupling through the coupling strength function
        $\Gamma_{ij}(\omega) = \delta_{ij} (v^2 \delta_{i1} + 0.5\cdot v^2 \delta_{i2})D(\omega)$, where 
        $D(\omega) = \left[(1+e^{-\nu(\omega-\omega_c)}) (1+e^{-\nu(\omega+\omega_c)}) \right]^{-1}$ describes a wide band with smooth cutoff.
        We set $\nu=10/v$ and $\omega_c=25v$.
        As $\Gamma_{ij}(\omega) \sim \delta_{ij}$ is diagonal in the orbital degree of freedom, the two orbitals are effectively decoupled and can in principle be studied independently of each other.
        We will exploit this fact to compare results from the InchMOSS scheme to results from previous formulations that were limited to single orbitals.
        The system is illustrated at the top of Fig.~\ref{fig:benchmark1}.
        For this benchmark, we consider the system at equilibrium, with inverse temperature $\beta = 0.1/v$.
        
        The bottom of Fig.~\ref{fig:benchmark1} shows the retarded GF of the first orbital (top panels) and the second orbitals (bottom panels) for different hybridization orders.
        The left panels showcase results obtained by treating the two orbitals separately, exploiting the fact that they decouple. Consequently, these results reflect a single-orbital perspective, illustrating the extent of previous methodological approaches.\cite{Erpenbeck_Quantum_2023}
        The right panels show the results from the multi-orbital scheme, which treats both orbitals in the system simultaneously (see yellow dashed line in the top of Fig.~\ref{fig:benchmark1}), and is agnostic to the fact that the two orbitals decouple.
        As orbital 1 is noninteracting, it is analytically solvable;\cite{Bruus_Many_2004, Haug_Quantum_2008} its analytic solution is shown as a red dashed line in each of the two top panels at the bottom of Fig.~\ref{fig:benchmark1}.
        For orbital two, which is interacting, there is no analytic result available. Here, we use the converged and numerically exact result from the single-orbital treatment (lower left panel) as the reference for the multi-orbital framework (red dashed line in lower right panel).
        
        The bottom of Fig.~\ref{fig:benchmark1} shows how the retarded GF converges to its respective reference with increasing inchworm order, for both the single-orbital and the multi-orbital treatment.
        In particular, the multi-orbital treatment recovers both the analytical reference for orbital 1 and the numerical reference for orbital 2 simultaneously, without taking advantage of the fact that the two orbitals are decoupled.
        The convergence behavior with respect to the inchworm order, however, is different for the separated single-orbital treatment and the multi-orbital treatment.
        While for the single-orbital treatment, the results converge at inchworm order $n\approx6$ for orbital one and order $n\approx2$ for orbital two, the results from the multi-orbital framework converge at inchworm order $n\approx12$ for both orbitals.
        This difference in convergence can be attributed to the fact that we limit the total number of hybridization lines per inchworm diagrams to $n$, rather than limiting the number of hybridization lines per orbital to $n$.
        The two-orbital expansion is therefore different at finite order, and contains fewer contributions.
        Work on the imaginary time multi-orbital algorithm\cite{Eidelstein_Multiorbital_2020} showed that this can be fully rectified by setting a maximum number of hybridizations per orbital as a numerical parameter instead, but this requires  implementing fast diagram summations that take advantage of the block structure.\cite{boag_inclusion-exclusion_2018}
        This optimization has not yet been introduced into the steady-state framework.

    \subsection{Mixing of orbitals through interactions and hybridizations}\label{sec:benchmarks_2}
    
        \begin{figure}
            \includegraphics{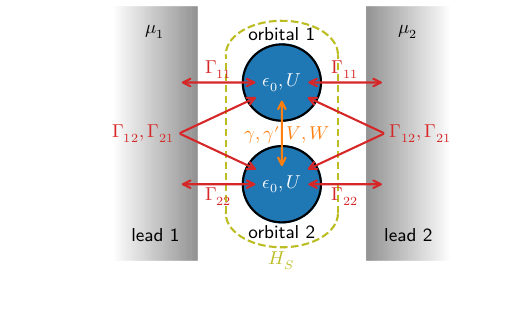}
            \includegraphics{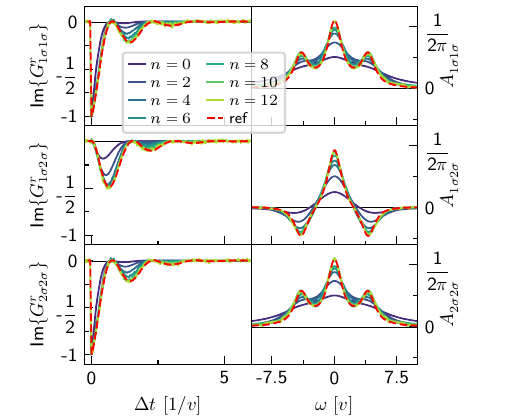}
            \caption{
                     Second benchmark for a two-orbital system.
                     Top: Sketch of the system. For high symmetry cases, this system can be decomposed into two decoupled orbitals. 
                     Bottom: Convergence of the retarded GF (left) and the spectral function (right) with inchworm order $n$.
                     The inverse temperature is $\beta=0.1/v$, the single-particle energies are $\epsilon_0 = -\gamma = -2v$, the two-particle interaction energies are $U = W = -V = -\gamma' = 2v$. 
                     The leads are described in the wide-band limit with a smooth cutoff with maximum coupling strength $\Gamma_{11} = \Gamma_{22} = \frac{3v^2}{4}$ and off-diagonal coupling strength 
                     $\Gamma_{12} = \Gamma_{21} = \frac{v^2}{4}$.
            }
            \label{fig:benchmark2}
        \end{figure}

        With our second benchmark, we establish the applicability of our methodology to general systems with intricate interactions between different orbitals and off-diagonal couplings to the environment.
        To this end, we consider a more general spinfull two-orbital impurity model,
        \begin{eqnarray}
            H_\text{S} &=&     \epsilon_0 \sum_{i\sigma} n_{i\sigma} 
                        + \gamma \sum_{\sigma} \left(d_{1\sigma}^\dagger d_{2\sigma} + d_{2\sigma}^\dagger d_{1\sigma}\right)
                        \label{eq:HS_benchmark2} \\ &&
                        + U \sum_{ij} n_{i\uparrow} n_{j\downarrow} 
                        + V \sum_{i\sigma} n_{i\sigma} \left( d_{1\overline\sigma}^\dagger d_{2\overline\sigma} + d_{2\overline\sigma}^\dagger d_{1\overline\sigma} \right)
                        \nonumber \\ &&
                        + W \sum_\sigma d_{1\sigma}^\dagger d_{1\overline\sigma} d_{2\overline\sigma}^\dagger d_{2\sigma} 
                        + \gamma' \left( d_{1\uparrow}^\dagger d_{1\downarrow}^\dagger d_{2\uparrow}  d_{2\downarrow} + \text{h.c.} \right) . \nonumber
        \end{eqnarray}
        Here, $\epsilon_0$ is the single-particle energy and $U$ is the Coulomb repulsion between electrons in different orbitals.
        $\gamma$ is the bare hopping between the two orbitals, $V$ is a population mediated hopping between the two orbitals, and $\gamma'$ is the pair-hopping between the two orbitals. 
        The spin-flip energy is $W$.
        As in the previous benchmark, we model the leads in the wide band with a smooth cutoff, $D(\omega) = \left[(1+e^{-\nu(\omega-\omega_c)}) (1+e^{-\nu(\omega+\omega_c)}) \right]^{-1}$, with $\nu=10/v$ and $\omega_c=25v$.
        The coupling strength function is chosen as
        $\Gamma_{ij}(\omega) = \left(\delta_{ij} 3v^2 + (1-\delta_{ij}) v^2\right) D(\omega)$, where $v$ is our unit of energy.
        An illustration of the system is shown at the top of Fig.~\ref{fig:benchmark2}, where we consider the environment to consist of two leads, for consistency with the setup of the previous benchmark.

        The system considered for this benchmark has different mechanisms that intertwine the two orbitals, either through interaction terms in the impurity as defined Eq.~(\ref{eq:HS_benchmark2}), or via the off-diagonal coupling to the leads.
        In order to benchmark this general scenario, we consider a high-symmetry case, which can be numerically diagonalized into two decoupled single-orbital problems.
        These can then be solved individually with the single-orbital steady-state inchworm method, providing a numerically exact reference value.
        Specifically, we consider the case that all interaction energies in the impurity are identical up to a sign, 
        with $U = W = \gamma = 2v$ and $\epsilon = V = \gamma' = -2v$.
        As before, the inverse temperature is $\beta=0.1/v$.

        The bottom of Fig.~\ref{fig:benchmark2} shows the retarded GF (left panels) and the associated spectral functions (right panels) for the general two-orbital impurity system.
        Due to the general setup, the GF and the spectral function not only have non-zero diagonal elements (top and bottom panels) associated with the two orbitals, but also non-zero off-diagonal elements (middle panels).
        We observe that the GF and the spectral function converge to the correct result with increasing inchworm order, whereby accurate results are obtained at order $n\approx10$.
        Note that the data presented here not only demonstrates that our methodology effectively treats general interacting multi-orbital systems, but also represents the first application of a real-time inchworm scheme to multi-orbital impurity systems.

\section{Conclusion}\label{sec:conclusion}

    In this work, we have developed a numerically exact CTQMC method for accurately characterizing quantum impurity models with multiple orbitals in the steady-state, both in and out of equilibrium. 
    The method, derived from the inchworm approach to quantum impurity models and based on the hybridization expansion, can account for arbitrary interactions between orbitals and general couplings between the impurity and its environment.
    Conceptually, the approach leverages the inchworm scheme's capability to overcome the multi-orbital sign problem,\cite{Eidelstein_Multiorbital_2020} a hallmark of general impurity-environment couplings, and exploits the time-translational invariance of the steady-state to directly compute steady-state properties without propagation.\cite{Erpenbeck_Quantum_2023} 
    We benchmarked the method against Anderson impurity models with two interacting spinfull orbitals, recovering analytical limits and results from previous methods. 
    Additionally, we assessed the numerical cost and its dependence on the number of impurity orbitals required for obtaining converged results.

    The method presented in this work not only enabled the first account of multi-orbital data obtained by a real-time inchworm scheme, but also lays the groundwork for the numerically exact treatment of a broad range of challenges.
    Aside from transport scenarios,
%     that were to a limited extent already considered in this work, 
    near term applications include quantum embedding theories like multi-orbital DMFT.
    Such theories often involve effective impurity models where hybridization with the environment mixes orbitals, giving rise to multi-orbital sign problems.
    Calculations with more than one or two orbitals remain rather prohibitively expensive at this time, but a few additional algorithmic improvements, as well as scaling up to leadership computing facilities, will soon make even larger multi-orbital systems possible to address.
    Furthermore, recent work has explored promising alternatives to Monte Carlo integration that may significantly cut down the computational expense, such as tensor cross-interpolation schemes\cite{nunez_learning_2022, erpenbeck_tensor_2023} and quasi-Monte Carlo algorithms.\cite{Macek_Quantum_2020, Bertrand_Quantum_2021, Strand_Inchworm_2023}
    \\

\section*{Acknowledgments}
    We thank Sergei Iskakov, Yang Yu, and Eitan Eidelstein for helpful discussions.
    Until August 31 2023, A.E. was funded by the Deutsche Forschungsgemeinschaft (DFG, German Research Foundation)—453644843. 
    E.G., T.B.,  W.-T.L., and A.E. (starting September 2023) were supported by the U.S. Department of Energy, Office of Science, Office of Advanced Scientific Computing Research and Office of Basic Energy Sciences, Scientific Discovery through Advanced Computing (SciDAC) program under Award No. DE-SC0022088. L.Z. was supported by the National Science Foundation under grant NSF QIS 2310182.
    G.C. was supported by the Israeli Science Foundation (Grant No. 2902/21) and by the PAZY foundation (Grant No. 318/78).
    This research used resources of the National Energy Research Scientific Computing Center, a DOE Office of Science User Facility supported by the Office of Science of the U.S. Department of Energy under Contract No. DE-AC02-05CH11231 using NERSC award BES-ERCAP0021805.
    The collaborative work between the US and Israel was enabled by NSF-BSF grant no. 2023720.

\section*{Data availability statement}
    The data that support the findings of this study are available from the corresponding author upon reasonable request.
    
\section*{Author Declarations}
    The authors have no conflicts of interest to disclose.

\appendix*
\section{Lowest order expressions of the inchworm expansion}\label{app:low_order}

    In the following, we outline how the lowest order inchworm expressions for the restricted propagators and the GF can be derived starting from their respective definitions in Eqs.~(\ref{eq:propagator_single}), (\ref{eq:propagator_double}), and (\ref{eq:GF_gtr_def}).\\
    
    \subsection{Single-branch propagator}\label{app:single_branch}
            We start by considering the restricted propagator on a single branch of the Keldysh-contour as defined in Eq.~(\ref{eq:propagator_single}).
            In order to derive explicit expressions for the inchworm expansion, which is equivalent to constructing the map $F_{\text{inch}}^\varphi$, we assume that the restricted single-branch propagator is known up to time $T<t$. 
            We rewrite Eq.~(\ref{eq:propagator_single_expansion}) by splitting all integrals into two parts at time $T$,
            \begin{widetext}
            \begin{eqnarray}
                \varphi_\alpha^\beta(t)
                &=&
                \sum_{n=0}^\infty
                \sum_{m=0}^n
                (-i)^n 
                            \int_T^t \hspace{-0.1cm} d\tau_1      \dots  \int_T^{\tau_{m-1}} \hspace{-0.5cm} d\tau_m
                            \times
                            \int_0^T \hspace{-0.2cm} d\tau_{m+1}  \dots  \int_0^{\tau_{n-1}} \hspace{-0.5cm} d\tau_n
                            \times
                            \nonumber
                            \\ && \times
                        \sum_\gamma
                        \text{Tr}_B \Big\lbrace \rho_B
                            \bra{\beta}
                            e^{-iH_0t}H_{\text{SB}}(\tau_1) \dots H_{\text{SB}}(\tau_m) e^{iH_0T}
                            \ket{\gamma}
                            \times
                            \bra{\gamma} 
                            e^{-iH_0T}
                            H_{\text{SB}}(\tau_{m+1}) \dots H_{\text{SB}}(\tau_n)\ket{\alpha}
                        \Big\rbrace
                        \label{eq:propagator_single_expansion_tinch}
                        .
            \end{eqnarray}
            This expression is formally equivalent to Eq.~(\ref{eq:propagator_single_expansion}), whereby $m$ enumerates the hybridization times $\tau_i>T$.
            Reordering the individual integrals and identifying restricted propagators at shorter times,  Eq.~(\ref{eq:propagator_single_expansion_tinch}) can be rewritten as 
            \begin{eqnarray}
                \varphi_\alpha^\beta(t)
                &=&
                \sum_\gamma
                        \text{Tr}_B \Big\lbrace \rho_B
                            \bra{\beta}
                            e^{-iH_0(t-T)}
                            \ket{\gamma}
                            \bra{\gamma} 
                            e^{-iH_0T}\ket{\alpha}
                        \Big\rbrace 
                        \nonumber
                        \\ &&
                    +
                        \int_T^t \hspace{-0.1cm} d\tau_1
                        \int_0^T \hspace{-0.2cm} d\tau_2
                        \sum_\gamma
                        \text{Tr}_B \Big\lbrace \rho_B
                            \bra{\beta}
                            e^{-iH_0t}H_{\text{SB}}(\tau_1)e^{iH_0T}
                            \ket{\gamma}
                            \times
                            \bra{\gamma} 
                            e^{-iH_0T}
                            H_{\text{SB}}(\tau_2)\ket{\alpha}
                        \Big\rbrace
                        + \dots
                    \\
                &=&
                    \sum_{\gamma}
                        \varphi_\gamma^\beta(t-T) \varphi_\alpha^\gamma(T) 
                    + 
                    \int_T^t \hspace{-0.1cm} d\tau_1
                    \int_0^T \hspace{-0.2cm} d\tau_2
                    \sum_{\gamma \nu_1\nu_2\nu_3\nu_4}
                        \varphi_{\nu_1}^\beta(t-\tau_1) 
                        \varphi_{\gamma}^{\nu_2}(\tau_1-T) 
                        \cdot
                        \Xi^{\nu_1 \nu_3}_{\nu_2 \nu_4}(\tau_1, \tau_2)
                        \cdot
                        \varphi_{\nu_3}^{\gamma}(T-\tau_2) 
                        \varphi_{\nu_4}^{\alpha}(\tau_2) 
                        + \dots
                        \ ,
                    \nonumber
                    \\ \label{eq:Feynman_inch_single_2}
            \end{eqnarray}
            \end{widetext}
            where we have introduced the shorthand
            \begin{eqnarray}
                \hspace{-0.5cm}
                \Xi^{\nu_1 \nu_3}_{\nu_2 \nu_4}(\tau_1, \tau_2) &=&
                        \text{Tr}_B \Big\lbrace 
                            \rho_B 
                            \bra{\nu_1} H_{\text{SB}}(\tau_1) \ket{\nu_2}
                            \bra{\nu_3} H_{\text{SB}}(\tau_2) \ket{\nu_4}
                        \Big\rbrace
                        \nonumber \\
                        &=& 
                        \sum_{ij\sigma} 
                        \Big(
                        \Delta_{ij\sigma}^<(\tau_1-\tau_2)
                        \bra{\nu_1} d_{i\sigma} \ket{\nu_2}
                        \bra{\nu_3} d_{j\sigma}^\dagger \ket{\nu_4}
                        \nonumber \\ &&
                        +
                        \Delta_{ji\sigma}^>(\tau_1-\tau_2)
                        \bra{\nu_1} d_{i\sigma}^\dagger \ket{\nu_2}
                        \bra{\nu_3} d_{j\sigma} \ket{\nu_4}
                        \Big)
                        ,
                        \nonumber \\
            \end{eqnarray}
            which encodes hybridizations between times larger and smaller than $T$.
            Eq.~(\ref{eq:Feynman_inch_single_2}) explicitly provides the first two lowest orders of the inchworm expansion, which incorporate restricted propagators over shorter time intervals. Consequently, even the second order inchworm expansion already accounts for a subset of the hybridization expansion up to infinite order.
            Note that we identify restricted propagators for time intervals both larger and smaller than $T$, which is possible due to the time-translational invariance of the steady-state that we are ultimately interested in. 
            In time-dependent scenarios, however, only restricted propagators at earlier time intervals are known, constraining the identification of restricted propagators to times strictly smaller than $T$. This scheme was proposed in the original work Ref.~\onlinecite{cohen_taming_2015}.\\

        \subsection{Two-branch propagator}\label{app:double_branch}
            Next, we consider the restricted propagator spanning both branches of the Keldysh-contour, which is defined in Eq.~(\ref{eq:propagator_double}).
            As the two-time propagator involves two time arguments, an inchworm expansion can be formulated for each of the time arguments. Here, we outline the derivation of the inchworm expansion for the second time argument, noting that deriving equations for the first time argument proceeds analogously.
            
            To derive explicit expressions for the inchworm expansion for the two-branch restricted propagator and establish the map $F_{\text{inch}}^\Phi$, we assume that $\Phi_\alpha^{\beta'\beta}(\tau',\tau)$ is known for any $\tau'\leq t'$ and $\tau\leq T$, i.e.\ the two-branch restricted propagator is only know up to time $T<t$ for the second time argument. 
            Expanding the two-branch restricted propagator in the impurity-bath coupling and splitting all integrals into two parts at time $T$, we can rewrite the two-branch restricted propagator as          
            \begin{widetext}
            \begin{eqnarray}
                \Phi_\alpha^{\beta'\beta}(t',t)
                &=&
                \sum_{n=0}^\infty
                \sum_{m=0}^n
                \sum_{k=0}^{n-m}
                (-i)^{m+k} (i)^{n-m-k}
                            \int_T^{t} \hspace{-0.1cm} d\tau_1      \dots  \int_T^{\tau_{m-1}} \hspace{-0.5cm} d\tau_m
                            \times
                            \int_0^T \hspace{-0.2cm} d\tau_{m+1}  \dots  \int_0^{\tau_{m+k-1}} \hspace{-0.5cm} d\tau_{m+k}
                            \times \int_0^{t'} \hspace{-0.2cm} d\tau_{m+k+1}  \dots  \int_0^{\tau_{n-1}} \hspace{-0.5cm} d\tau_{n}
                            \times
                            \nonumber
                            \\ && \times
                        \sum_\gamma
                        \text{Tr}_B \Big\lbrace \rho_B
                            \bra{\alpha}e^{iHt}
                            H_{\text{SB}}(\tau_{n}) \dots H_{\text{SB}}(\tau_{m+k+1})
                            e^{iH_0t'}
                            \ket{\beta'}
                            \times \label{eq:propagator_double_expansion_tinch}\\ &&
                            \hspace*{1.5cm}
                            \times
                            \bra{\beta}
                            e^{-iH_0t}H_{\text{SB}}(\tau_1) \dots H_{\text{SB}}(\tau_m) e^{iH_0T}
                            \ket{\gamma}
                            \times
                            \bra{\gamma} 
                            e^{-iH_0T}
                            H_{\text{SB}}(\tau_{m+1}) \dots H_{\text{SB}}(\tau_{m+k})\ket{\alpha}
                        \Big\rbrace
                        \nonumber
                        .
            \end{eqnarray}
            This expression is formally exact, and as before, $m$ enumerates the hybridization times $\tau_i>T$, while $k$ specifies the number of hybridization times on the two branches.
            Reordering the individual integrals and identifying singe-branch and two-branch restricted propagators at shorter time intervals, Eq.~(\ref{eq:propagator_double_expansion_tinch}) can be rewritten as 
            \begin{eqnarray}
                \Phi_\alpha^{\beta'\beta}(t',t) 
                &=&	
                \sum_\gamma
                        \text{Tr}_\text{B} \left\lbrace \rho_\text{B}
								\bra{\alpha}e^{iHt'}\ket{\beta'} 
								\times
								\bra{\beta}e^{-iH_0(t-T)}
                                \ket{\gamma} \bra{\gamma} 
                                e^{-iH_0T}\ket{\alpha}
								\right\rbrace 
                        \nonumber
                        \\ &&
                    +
                        \int_T^{t} \hspace{-0.1cm} d\tau_1
                        \int_0^T \hspace{-0.2cm} d\tau_2
                        \sum_\gamma
                        \text{Tr}_B \Big\lbrace \rho_B
                            \bra{\alpha}e^{iHt'}\ket{\beta'} 
                            \bra{\beta}
                            e^{-iH_0t}H_{\text{SB}}(\tau_1)e^{iH_0T}
                            \ket{\gamma}
                            \times
                            \bra{\gamma} 
                            e^{-iH_0T}
                            H_{\text{SB}}(\tau_2)\ket{\alpha}
                        \Big\rbrace
                        \nonumber
                        \\ &&
                    -
                        \int_T^{t}  \hspace{-0.1cm} d\tau_1
                        \int_0^{t'} \hspace{-0.2cm} d\tau_2
                        \sum_\gamma
                        \text{Tr}_B \Big\lbrace \rho_B
                            \bra{\alpha}H_{\text{SB}}(\tau_2)e^{iHt'}\ket{\beta'} 
                            \bra{\beta}
                            e^{-iH_0t}H_{\text{SB}}(\tau_1)e^{iH_0T}
                            \ket{\gamma}
                            \times
                            \bra{\gamma} 
                            e^{-iH_0T}
                            \ket{\alpha}
                        \Big\rbrace
                        + \dots
                    \\
                &=&
                    \sum_{\gamma}
                        \varphi_\gamma^{\beta}(t-T) \Phi_\alpha^{\beta'\gamma}(t',T)
                    \\&&
                    + 
                    \int_T^{t} \hspace{-0.1cm} d\tau_1
                    \int_0^T \hspace{-0.2cm} d\tau_2
                    \sum_{\gamma \nu_1\nu_2\nu_3\nu_4}
                        \varphi_{\nu_1}^{\beta}(t-\tau_1) 
                        \varphi_{\gamma}^{\nu_2}(\tau_1-T) 
                        \cdot
                        \Xi^{\nu_1 \nu_3}_{\nu_2 \nu_4}(\tau_1, \tau_2)
                        \cdot
                        \varphi_{\nu_3}^{\gamma}(T-\tau_2) 
                        \Phi_\alpha^{\beta'\nu_4}(t', \tau_2)
                    \nonumber \\ &&
                    -
                    \int_T^{t}  \hspace{-0.1cm} d\tau_1
                    \int_0^{t'} \hspace{-0.2cm} d\tau_2
                    \sum_{\gamma \nu_1\nu_2\nu_3\nu_4}
                        \varphi_{\nu_1}^{\beta}(t-\tau_1) 
                        \varphi_{\gamma}^{\nu_2}(\tau_1-T) 
                        \cdot
                        \Xi^{\nu_1 \nu_3}_{\nu_2 \nu_4}(\tau_1, \tau_2)
                        \cdot
                        \Phi_\alpha^{\nu_4\gamma}(\tau_2, T)
                        \varphi_{\nu_3}^{\beta'}(\tau_2-t') 
                        + \dots
                        \ .
                    \label{eq:Feynman_inch_double_2}
            \end{eqnarray}
            \end{widetext}
            Eq.~(\ref{eq:Feynman_inch_double_2}) explicitly provides the first two lowest orders of the inchworm expansion, which incorporate single-branch and two-branch restricted propagators at previous time intervals. 
            Again, this means that the second order inchworm expansion already incorporates a subset of the hybridization expansion up to infinite order.
            Note that Eq.~(\ref{eq:Feynman_inch_double_2}) contains two distinct terms at the second order of the inchworm expansion. These terms distinguish between hybridization events confined to a single Keldysh branch, and those bridging the two branches.
            The equations for higher-order contributions are more involved, as it necessitates accounting for all hybridization events and their positions on the different Keldysh branches.
            The Feynman diagrams corresponding to the three terms given Eq.~(\ref{eq:Feynman_inch_double_2}) are visualized in the first three lines on the right hand side of Fig.~\ref{fig:Feynman_inch_double}.

        \subsection{Green's functions}\label{app:GF}
            Finally, we provide the expressions for the GF in terms of the single- and two-branch restricted propagators.
            We explicitly consider the greater GF as defined in Eq.~(\ref{eq:GF_gtr_def}), and note that the derivation for the lesser GF is equivalent but $t$ and $t'$ as well as fermionic creation and annihilation operators are exchanged.
            
            Starting from its definition, we rewrite the greater GF as
            \begin{widetext}
            \begin{eqnarray}
                G^>_{i\sigma j\sigma'}(t, t') &=&   -i \braket{ d_{i\sigma}(t) d^\dagger_{j\sigma'}(t') } \\
                &=&
                    \sum_{\beta\gamma\delta\epsilon}
                        \text{Tr}_\text{B} \left\lbrace \rho_\text{B}
								\bra{\alpha}e^{iHt}\ket{\beta} 
								\bra{\beta} d_{i\sigma} \ket{\gamma} 
								\bra{\gamma} e^{-iH(t-t')} \ket{\delta} 
								\bra{\delta} d^\dagger_{j\sigma'} \ket{\epsilon} 
                                \bra{\epsilon} e^{-iHt'}\ket{\alpha}
								\right\rbrace 
                        \label{eq:GF_gtr_time_dep}
                        \\
                &=&
                    \sum_{\beta\gamma\delta\epsilon}
								\bra{\beta} d_{i\sigma} \ket{\gamma} 
								\varphi_\delta^\gamma(t-t')
								\bra{\delta} d^\dagger_{j\sigma'} \ket{\epsilon} 
                                \Phi_\alpha^{\beta\epsilon}(t, t')
                    \nonumber \\ &&
                    + 
                    \sum_{\beta\gamma\delta\epsilon}
                    \sum_{\nu_1\nu_2\nu_3\nu_4}
                    \int_{t'}^{t} \hspace{-0.1cm} d\tau_1
                    \int_0^{t'} \hspace{-0.2cm} d\tau_2
                    \bra{\beta} d_{i\sigma} \ket{\gamma} 
								\varphi_{\nu_1}^\gamma(t-\tau_1)
								\varphi_\delta^{\nu_2}(\tau_1-t')
								\bra{\delta} d^\dagger_{j\sigma'} \ket{\epsilon} 
								\varphi_{\nu_3}^\epsilon(t'-\tau_2)
                                \Phi_\alpha^{\beta{\nu_4}}(t, \tau_2)
                                \Xi^{\nu_1 \nu_3}_{\nu_2 \nu_4}(\tau_1, \tau_2)
                    \nonumber \\&&
                    -
                    \sum_{\beta\gamma\delta\epsilon}
                    \sum_{\nu_1\nu_2\nu_3\nu_4}
                    \int_{t'}^{t} \hspace{-0.1cm} d\tau_1
                    \int_0^{t} \hspace{-0.2cm} d\tau_2
                    \bra{\beta} d_{i\sigma} \ket{\gamma} 
								\varphi_{\nu_1}^\gamma(t-\tau_1)
								\varphi_\delta^{\nu_2}(\tau_1-t')
								\bra{\delta} d^\dagger_{j\sigma'} \ket{\epsilon} 
                                \Phi_\alpha^{\nu_4\epsilon}(\tau_2, t')
                                \varphi_{\nu_3}^\beta(\tau_2-t)
                                \Xi^{\nu_1 \nu_3}_{\nu_2 \nu_4}(\tau_1, \tau_2)
                    \nonumber \\ &&
                    + \dots \ ,
                    \label{eq:Feynman_inch_GF}
            \end{eqnarray}
            \end{widetext}
            where we have inserted the time-dependence of the creation and annihilation operators in Eq.~(\ref{eq:GF_gtr_time_dep}). 
            This time-dependence is then subjected to the hybridization expansion, i.e.\ the exponential functions are expanded in the impurity-bath coupling.
            Regrouping the individual terms of the hybridization expansion, the GF can be expressed in terms of the restricted propagators of the inchworm methodology, see Eq.~(\ref{eq:Feynman_inch_GF}). 
            As the restricted propagators already contain contributions to the hybridization expansion up to arbitrary order, even the first term in Eq.~(\ref{eq:Feynman_inch_GF}), which is a product of a single-branch and a two-branch restricted propagator, already accounts for an infinite subset of all contributions to the hybridization expansion.
            Moreover, we point out the similarity between the expressions for the two-time restricted propagator and for the GF in Eqs.~(\ref{eq:Feynman_inch_double_2}) and (\ref{eq:Feynman_inch_GF}). 
            This allows us to evaluate higher-order contributions to the GF using inchworm proper diagrammatics, as is visualized by Feynman diagrams in Fig.~\ref{fig:Feynman_inch_GF}.

\bibliography{bib.bib}

\end{document}